\documentclass[%
 reprint,
 amsmath,amssymb,
 aps,
prl
]{revtex4-2}
\usepackage{float}
\usepackage{graphicx}
\usepackage{dcolumn}
\usepackage{bm}

\usepackage{orcidlink}
\usepackage{hyperref}
\hypersetup{colorlinks=true, citecolor=blue, linkcolor = magenta}
\usepackage{footmisc}
\begin{document}

\preprint{APS/123-QED}

\title{Stable colored black holes with quartic self-interactions}

\author{Jos\'e~F.~Rodr\'iguez-Ruiz \orcidlink{0000-0003-3627-5084}} 
\email{jrodriguez154@uan.edu.co}
\affiliation{Departamento de F\'isica, Universidad Antonio Nari\~no, Cra 3 Este \# 47A - 15, Bogot\'a D.C. 110231, Colombia}

\author{Gabriel~G\'omez \orcidlink{0000-0002-3618-9824}}
\email{luis.gomezd@umayor.cl}
\affiliation{Centro Multidisciplinario de F\'isica, Vicerrector\'ia de Investigaci\'on, Universidad Mayor,\\ Camino La Pir\'amide 5750,  Huechuraba, 8580745, Santiago, Chile}
\date{\today}

\begin{abstract}

We analytically prove the linear radial stability of non-Abelian black holes with quartic self-interactions. The background, constructed from the Wu-Yang magnetic monopole ansatz, is an exact black-hole solution carrying a non-Abelian magnetic charge $Q_{\rm NA}^2$ controlled by a single coupling parameter $\chi$, and admits two distinct branches. 
The odd sector is always stable, while in the even sector the effective potential is positive for branch~I and negative for branch~II, establishing stability and potential instability, respectively. The potential instability of branch~II is consistent with its connection to the perturbatively unstable Einstein-Yang-Mills Reissner-Nordstr\"{o}m solution. Branch~I remains linearly stable throughout the physical domain of $\chi$ where the solutions are regular and free of naked singularities. The stability of branch~I is further confirmed through a numerical computation of the quasinormal mode spectrum using Leaver's continued fraction method and time-domain evolution supplemented with Prony method, showing no unstable modes. Our results prove the existence of the first linearly stable asymptotically flat hairy black holes in four dimensions with a minimally coupled non-Abelian Proca self-interaction.

\end{abstract}

\keywords{Black holes}
\maketitle

\paragraph*{Introduction.}
The no-hair theorem establishes that stationary black holes (BHs) in general relativity (GR) coupled to standard matter are fully characterized by mass, (Abelian) charge, and angular momentum~\cite{Israel:1967wq,Carter:1971zc, Robinson:1975bv}.  Whether GR coupled to non-Abelian gauge fields admits stable hairy BH solutions has remained an open question for three decades~\cite{Volkov:1998cc}. To date, no non-Abelian hairy BH stable against linear radial perturbations has been found in asymptotically flat, $d=4$ GR with minimally coupled fields. In this Letter, we show that this situation changes once an additional quartic self-interaction is introduced in the vector sector, yielding the first non-Abelian BH that is linearly stable against radial perturbations. 
 
The colored BH solutions of the Einstein-Yang-Mills (EYM) theory~\cite{Bizon:1990sr,Volkov:1989fi} provided the first counterexamples to no-hair expectations in GR, but were promptly shown to be perturbatively unstable~\cite{Lee:1991qs,Straumann:1990as}. In particular, the EYM Reissner-Nordstr\"{o}m (RN) solution is unstable~\cite{Lee:1991qs}. All subsequent attempts to stabilize non-Abelian hair in asymptotically flat $d=4$ spacetime have required departures from the minimal framework: either a negative cosmological constant~\cite{Winstanley:1998sn} or higher-order gauge field terms~\cite{Radu:2011ip} that lack natural dimensional or topological motivation in four dimensions. In addition, when the local symmetry is broken, asymptotically flat non-Abelian Proca BHs were found in~\cite{Greene:1992fw}, but unfortunately, they are radially unstable, see \cite{ Ponglertsakul:2016fxj} and references therein.

We use geometric units, setting $c = G = 1$, where $c$ denotes the speed of light and $G$ represents the universal gravitational constant. In this Letter we consider the classical regime of BHs. Greek indices correspond to space-time coordinates, ranging from 0 to 3, while Latin indices denote SU(2) group indices, varying from 1 to 3.

\textit{The model.} We consider an $\mathrm{SU}(2)$ vector field $\mathbf{B}$ minimally coupled to gravity, described by the action~\cite{Gomez:2023wei}
\begin{equation}
    S = \frac{1}{16\pi}\int\sqrt{-g}\,d^4x\bigl[R - F_{a\mu\nu}F^{a\mu\nu}
    +\chi\, B_{a\mu}B^a{}_\nu B_b{}^\mu B^{b\nu}\bigr],
    \label{eq:action}
\end{equation}
where $F_{a\mu\nu}$ is the SU(2) field strength and $\chi$ is a coupling constant with dimensions of inverse length squared. We use the gauge coupling constant $\tilde{g}$, which has units of inverse length, to define the normalized variables, $r\to r\tilde{g}$, $M\to M\tilde{g}$, $\chi\to\chi/\tilde{g}^2$, $Q_{\rm NA}\to Q_{\rm NA}\tilde{g}$. The form of the equations in the normalized variables can be obtained effectively by setting $\tilde{g}=1$. Without loss of generality, we further set $M = 1$, since any physical BH mass can be accommodated by an appropriate choice of $\tilde{g}$  \footnote{For instance, if we consider Sgr A*, the gauge coupling  in geometrized units is $\tilde{g}\sim 1.6\times10^{-10}$ m$^{-1}$. Physical lengths can be obtained as $L_{\rm phys} = L_{\rm norm }/\tilde{g}$.}. 

The quartic self-interaction term in Eq.~\eqref{eq:action} arises naturally as one of the Lagrangian building blocks of the Generalized SU(2) Proca (GSU2P) theory~\cite{Allys:2016kbq,GallegoCadavid:2020dho}, a ghost-free vector-tensor theory constructed in the same spirit of Horndeski~\cite{Horndeski:1974wa}, but the action~\eqref{eq:action} stands on its own as a well-defined theory of GR with non-Abelian self-interacting vector fields.
 
Exact analytical non-Abelian RN BH solutions of this theory were found in Ref.~\cite{Gomez:2023wei}, arising from the Wu-Yang magnetic monopole ansatz. The solutions are characterized by an effective non-Abelian magnetic charge
\begin{equation}
    Q^2_{\rm NA\, I,II} = \frac{1-4\chi(5+2\chi)\mp(1+8\chi)^{3/2}}{2(1-\chi)^3},
\end{equation}
with two branches labeled I ($-$ sign) and II ($+$ sign), and a
vector field amplitude $w_0 = w_{\rm I,II}$ given by
\begin{equation}
    w_0 = w_{\rm I,II} = \frac{1+2\chi\pm\sqrt{1+8\chi}}{2-2\chi}.
\end{equation} 
 
Branch~I corresponds to the physically nontrivial domain $\chi \in [\chi_{\rm ext}, 0]$, with $\chi_{\rm ext} \approx -0.0902$.  In this interval $Q^2_{\rm NA,I} \in [0, 1]$ is real and positive, $w_{\rm I}$ remains finite and regular, and no naked singularity forms.  The solution interpolates continuously between the extremal RN BH at $\chi = \chi_{\rm ext}$ and the Schwarzschild solution at $\chi = 0$. Branch~II corresponds to $\chi \in [0, \infty)$, where $Q^2_{\rm NA,II} \in [0, 1]$ is likewise real and positive. This branch interpolates between the EYM RN solution at $\chi = 0$ and the Schwarzschild solution as $\chi \to \infty$.
A detailed study of the astrophysical implications of these solutions, including Event Horizon Telescope shadow constraints from Sagittarius~A$^*$, critical accretion rates, and innermost stable circular orbits, was presented in Ref.~\cite{Gomez:2023wei}, where perturbative stability was identified as an open problem.
 
\paragraph*{Perturbations.}
The line element of the spherically symmetric spacetime in Schwarzschild coordinates is
\begin{equation}
    ds^2 = -e^{-2\delta(t,r)}f(t,r)dt^2 + f(t,r)^{-1}dr^2 +r^2d\Omega^2.
\end{equation}
The most general spherically symmetric configuration for the vector fields is described by \cite{Witten:1976ck}

\begin{multline}
    \mathbf{B} = \tau^a\Biggl[A_0(t,r) \frac{x_a}{r}dt + A_1(t,r) \frac{x_a x_j}{r^2} dx^j
        \\+ \frac{A_2(t,r)}{r}\biggl(\delta_{aj} - \frac{x_a x_j}{r^2}\biggr) dx^j  -\epsilon_{ajk}x^j\frac{[1+w(t,r)]}{r^2}dx^k\Biggr] \,,  
\end{multline}
where $\tau_i = -i\sigma_i/2$ is the anti-Hermitian basis for the SU(2) algebra and $\sigma_i$ are the Pauli matrices. The explicit form of the Einstein field equations and the field equations for $\mathbf{B}$ in this spherically symmetric configuration can be found in \footnote{Supplemental Material at [url] for a brief discussion of the full field equations and the numerical methods used to find the perturbation spectrum, which includes Ref. \cite{xAct} }\edef\suppfootnum{\thefootnote}.

The static equilibrium configuration obtained in Ref.~\cite{Gomez:2023wei} corresponds to a purely magnetic solution, for which \(A_0=A_1=A_2=0\) and \(w=w_0\). In the Wu-Yang configuration, the background metric functions are
\begin{equation}
    f_0(r)
    =
    1-\frac{2}{r}+\frac{Q_{\rm NA}^2}{r^2},
    \qquad
    \delta_0(r)=0.
\end{equation}
with outer and inner horizons located at $r_+$ and $r_-$, respectively.

\begin{figure}[!t]
    \centering
    \includegraphics[width=\linewidth]{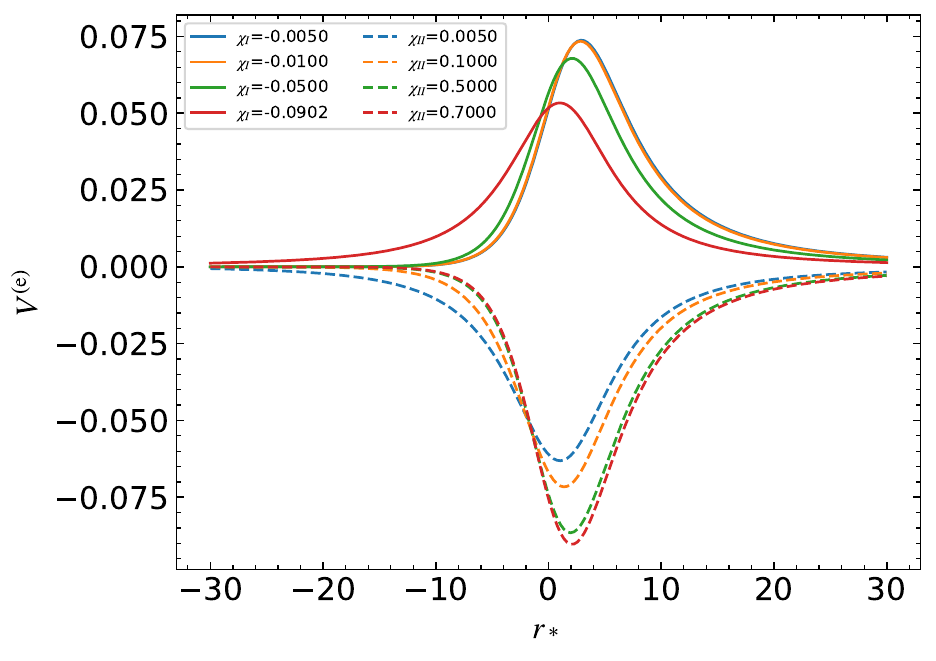}
    \caption{Effective potential for even-parity perturbations, given in \eqref{eq:Ve}. The potential is strictly positive for branch~I and strictly negative for branch~II. 
    }
    \label{fig:Veven}
\end{figure}

\begin{figure}[!t]
    \centering
    \includegraphics[width=\linewidth]{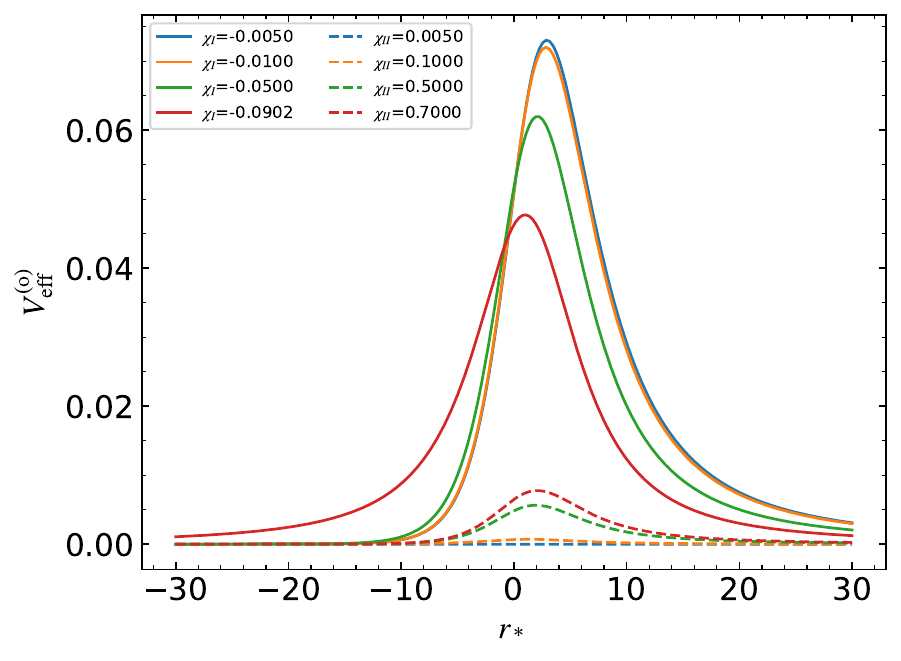}
    \caption{Effective potential for odd perturbations given by \eqref{eq:Veffo}. The potential is always positive. 
    }
    \label{fig:Vodd}
\end{figure}

Linearizing around this background, we introduce the following perturbations:
\begin{align}
    f(t,r) &= f_0(r) + \epsilon f_1(t,r),\\
    \delta(t,r) &= \epsilon \delta_1(t,r),\\
    A_0(t,r) &= \epsilon \delta A_0(t,r) \equiv \epsilon u(t,r),\\
    A_1(t,r) &= \epsilon \delta A_1(t,r) \equiv \epsilon v(t,r),\\
    A_2(t,r) &= \epsilon \delta A_2(t,r) \equiv \epsilon \tilde{w}(t,r),\\
    w (t,r) &= w_{0} + \epsilon w_1(t,r).
\end{align} 
Since the background is invariant under rotations, the perturbed equations decouple into two sectors according to their behavior under parity transformations.
\subparagraph*{Even perturbations.}
At linear order in $\epsilon$, the even sector is governed by
\begin{equation}
    \partial_r\delta_1=0,
\end{equation}
\begin{equation}
    \frac{\partial_t f_1}{rf_0}  = 0,
    \label{eqn:chi1-f1}
\end{equation}
\begin{equation}
      \frac{\partial_r f_1}{r}+\frac{f_1}{r^2}+ \left[\frac{4 w_0 \left(w_0^2-1\right)}{r^4}-\frac{4 \chi  (w_0+1)^3}{r^4}\right]w_1=0, \label{eqn:f1prime}
\end{equation}
\begin{multline}
    -f_0' \partial_r w_1-f_0 \partial_r^2  w_1+\frac{\partial_t^2w_1}{f_0}\\
    -\frac{w_1 \left[3 \chi +3 (\chi -1) w_0^2+6 \chi  w_0+1\right]}{r^2}=0,\label{eqn:w1original}
\end{multline}
where prime denotes the total radial derivative. The metric perturbations are non-dynamical: the coefficient of $w_1$ in \eqref{eqn:f1prime} vanishes identically on solutions of the background field equation.
Now, we define the tortoise coordinate $r_*$ via $dr_*/dr = 1/f_0$. Assuming harmonic time dependence $w_1 = e^{-i\omega t}w_1(r)$, the perturbation equation reduces to the Schr\"{o}dinger-like form
\begin{equation}
     - \frac{d^2 w_1}{d r_*^2}+ V^{(e)} w_1 = \omega^2 w_1,
\end{equation}
with effective potential
\begin{equation}
    V^{({\rm e})} = f_0\,\frac{P(w_0,\chi)}{r^2},\qquad
    P \equiv 3w_0^2 - 3\chi(1+w_0)^2 - 1,
    \label{eq:Ve}
\end{equation}
which vanishes at both the horizon and spatial infinity, i.e.,
$V^{\rm (e)}\to0$ as $r\to r_+$ and $V^{\rm (e)}\to0$ as $r\to \infty$. 
Since $f_0 \geq 0$ for $r \geq r_+$, the sign of $V^{(\rm e)}$ is determined entirely by $P(w_0,\chi)$. Introducing
$s \equiv \sqrt{1+8\chi}$ 
with $s \in [s_{\rm ext},1]$ where $s_{\rm ext}\approx0.53$, for branch I and $s>1$ for branch II, the two branches simplify to
\begin{equation}
    w_{\rm I} = \frac{s+1}{3-s},
    \qquad
    w_{\rm II} = \frac{1-s}{s+3}.
\end{equation}  
Direct substitution into $P$ yields the closed-form expressions
\begin{align}
    P_{\rm I}  &= \frac{4s}{3-s},
    \label{eq:PI}\\[4pt]
    P_{\rm II} &= -\frac{4s}{3+s}.
    \label{eq:PII}
\end{align}
Branch~I: Since $s \in [s_{\rm ext},1] \subset (0,3)$, both the numerator and the denominator of $P_{\rm I}$ in Eq.~\eqref{eq:PI} are strictly positive.  Hence $V^{(\rm e)}_{\rm I} = f_0 P_{\rm I}/r^2 \geq 0$ throughout the physical parameter space, with equality attained only at the horizon $r=r_+$ and asymptotically as $r\to\infty$. Branch II: Since $s>1$, $P_{\rm II} < 0$. Therefore, $V_{\rm II}^{(e)} \leq 0$ for all $r \geq r_+$ and all $\chi > 0$.
The potential $V^{\rm (e)}$ for selected values of $\chi$ is shown in Fig. \ref{fig:Veven}.
\subparagraph*{Odd perturbations.}
For this sector, the perturbations are
\begin{multline}
-\frac{2 w_0 \partial_t\tilde{w}+f_0 r \left(2\partial_t v+r \partial_{tr}^2v\right)+2 u w_0^2}{r^2}\\
+\frac{2 f_0 \partial_r u}{r}+f_0 \partial_r^2 u= 0, \label{eqn:u} 
\end{multline}
\begin{equation}
    \frac{ \partial^2_{tr}u}{f_0}-\frac{ \partial_t^2 v}{f_0}-\frac{2 w_0^2 v +2 w_0\partial_r\tilde{w} }{r^2}=0, \label{eqn:v}
\end{equation}
\begin{multline}
    -f_0 \left(\partial_r^2\tilde{w}+w_0 \partial_r v\right)+\frac{\partial_t^2\tilde{w}}{f_0}-f_0' \partial_r \tilde{w}\\-\frac{\tilde{w} \left[w_0 \left(\chi w_0-w_0+2 \chi\right)+\chi+1\right]}{r^2}\\+\frac{ w_0 \partial_t u}{f_0}-w_0 f_0'v = 0. \label{eqn:wtilde}
\end{multline}
We now impose the generalized ``Lorentz gauge'',  which is a consequence  of the field equations of $\mathbf{B}$ and it is not a choice. This constraint is given by  \cite{Gomez:2025pag}
\begin{equation}
    \nabla^\alpha(\chi B_{a\mu}B_{b}{}^{\mu}B^b_\alpha)=0.
\end{equation}
At linear order, this reduces to
\begin{equation}
 \frac{2 \chi\left(w_0+1\right)^2 \tilde{w}}{r^4} =0.
\end{equation}
Notably, this condition does not involve the perturbation $w_1$, but depends explicitly on the coupling $\chi$. Since $w_0 \neq-1$ for both branches \citep[see Fig. 1 in Ref.][]{Gomez:2023wei} and considering $\chi \neq 0$, we obtain $\tilde{w}=0$. This result simplifies the equations. Now, Eq.~\eqref{eqn:wtilde} becomes a constraint
\begin{equation}
    \partial_tu = f_0 \left(f_0 \partial_r v+v f_0'\right),
\end{equation}
which can be used to eliminate $u$ from \eqref{eqn:v}, leading to  the master equation
\begin{equation}
    f_0\partial_t^2 v-\partial_r\left(f_0^3\partial_rv\right) + V^{(o)}v= 0,
\end{equation}
where
\begin{equation}
    V^{\rm (o)} = f_0\left[\frac{f_0 \left(2 w_0^2-r^2 f_0''\right)-r^2f_0'^2}{r^2}\right],
\end{equation}
is the potential governing the odd perturbations. Again, we use the tortoise coordinate $r_*$, but we define the new field variable $\psi\equiv vf_0$. For periodic perturbations $\psi = e^{-i\omega t}\psi(r)$, the master equation again takes a Schr\"{o}dinger-like form 
\begin{equation}
    -\frac{d^2\psi}{dr_*^2} + V_{\mathrm{eff}}^{\rm(o)}\,\psi = \omega^2\psi,
    \label{eq:schrodinger}
\end{equation}
with the effective potential
\begin{equation}
    V_{\mathrm{eff}}^{(\rm o)}= \frac{2w_0^2f_0}{r^2} .
    \label{eq:Veffo}
\end{equation}
This potential is always manifestly positive. Figure~\ref{fig:Vodd} illustrates how the coupling $\chi$ modifies its profile.

\subparagraph*{Stability Analysis.}
Following the linear stability analysis of  Wald~\cite{Wald:1979lth}, the operator $\mathcal{A} = -d^2/d{r_*}^2 + V
$ is essentially self-adjoint on the appropriate domain of square-integrable functions $L^2(r_*)$. Since $V \geq 0$ for both sectors of branch~I, the operator is non-negative definite and its spectrum lies in $[0, \infty)$, implying that it is bounded below by zero. By Wald's argument, no exponentially growing normalizable mode can therefore exist, establishing the linear stability of branch I. 

Quasinormal modes (QNM) are defined by purely ingoing radiation at the horizon and purely outgoing radiation at spatial infinity, respectively. We numerically confirmed the stability by calculating the QNM spectrum using two different methods: Leaver's continued-fraction method \cite{Leaver:1985ax, Leaver:1990zz} and leapfrog integration in the time domain at a fixed value of the radial coordinate, followed by extraction of the frequencies with the Prony method. We show the fundamental mode, i.e., the least damped mode, in Fig. \ref{fig:spectrum}. Details on both methods can be found in the Supplemental Material \footnotemark[\suppfootnum].

\begin{figure}
    \centering
    \includegraphics[width=\linewidth]{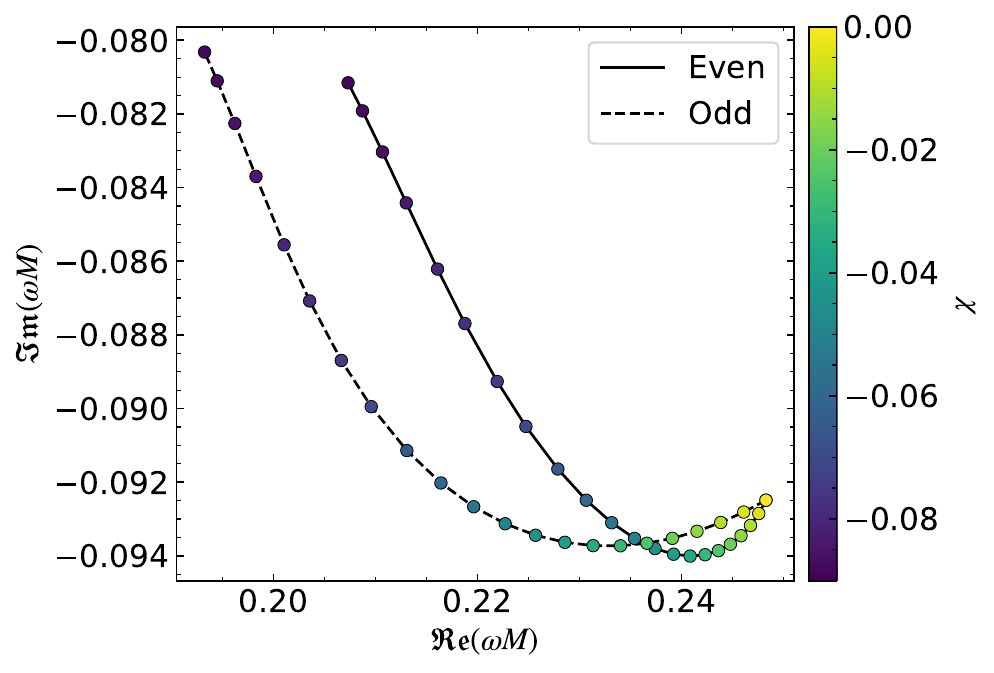}
    \caption{QNM spectrum of branch~I obtained using Leaver’s continued-fraction method. The figure displays the fundamental mode as a function of the coupling parameter $\chi$. The negative imaginary part of the eigenfrequency confirms that the fundamental mode is damped, providing further evidence for the stability within the explored parameter range.}
    \label{fig:spectrum}
\end{figure}
The QNM appear singular at $r_+$ and at infinity in
Schwarzschild coordinates, but this is a coordinate artifact.
Introducing advanced and retarded Eddington-Finkelstein coordinates $V=t+r_*$ and $U=t-r_*$,
the perturbation for $w_1$ or $\psi$ factorises as $\Phi = e^{-i\omega V}\,
\widetilde{\Phi}_V(r)$ with $\widetilde{\Phi}_V \equiv e^{+i\omega
r_*}\Phi$ near the horizon, and as $\Phi = e^{-i\omega U}\,
\widetilde{\Phi}_U(r)$ with $\widetilde{\Phi}_U \equiv e^{-i\omega
r_*}\Phi$ near infinity.  The asymptotic solutions give
$\widetilde{\Phi}_V(r_+) = 1 + O(r-r_+)$ and
$|\widetilde{\Phi}_U(r)| = 1 + O(r^{-1})$, both finite.
Consequently, the perturbations of the invariant observables $F^2 = F_{a\mu\nu}F^{a\mu\nu}$ and $\rho$, energy density associated with $\mathbf{B}$, given by
\begin{align}
    \delta F^2 &= \frac{8w_0(w_0^2-1)\,w_1}{r^4},\\[4pt]
    \delta\rho &= \frac{(w_0^2-1)^2 f_1 + \chi(w_0+1)^4 f_1
                       + O(f_0)}{8\pi r^4},
\end{align}
are finite at $r_+$ (the $O(f_0)$ terms vanish there) and decay as
$r^{-4}$ at infinity.  
Hence, the total perturbed energy and fluxes are both finite, confirming the physical consistency of the stability analysis.
\paragraph*{Discussion and outlook.} 
We have analytically proved that the non-Abelian RN BH solutions of branch~I are linearly stable against radial perturbations throughout the entire physical parameter space $\chi\in[\chi_{\rm ext},0]$, with $\chi_{\rm ext}\simeq-0.0902$, in the presence of quartic self-interactions minimally coupled to gravity.
The sign of $\chi$ corresponds to a repulsive quartic self-interaction that reinforces the YM contribution and indeed stabilizes the hair.  The proof is exact and requires neither approximations nor fine-tuning: the odd sector stability follows from an algebraic cancelation that yields $V_{\rm eff}^{({\rm o})}=2w_0^2f_0/r^2\geq 0$, and the even sector stability follows from the closed-form result $P_{\rm I}=4s/(3-s)>0$.  Branch~II exhibits a negative  potential in the even sector, signaling instability, consistent with its connection at $\chi=0$ to the unstable
EYM RN solution~\cite{Lee:1991qs}.
Furthermore, we have computed the QNM spectrum of branch~I and verified that the fundamental mode has a negative imaginary part of the eigenfrequency throughout the explored parameter range, providing an independent numerical confirmation of the linear stability established by the analytical analysis. The two sectors are not isospectral.

It is important to assess whether the stable Branch I is compatible with the observational constraints derived from observations of Sgr A$^{\star}$. In Ref. \cite{Gomez:2023wei}, we constrained the model parameters $\tilde{g}$ and $\chi$ using the VLTI and Keck measurements of the Sgr A$^{\star}$ shadow, with the resulting constraints summarized in Fig. 3 therein. In the normalization adopted here, $\tilde{g}M=1$, the observational constraints can be expressed in terms of the combination $\chi/\tilde{g}^2$. The VLTI observations require $-0.067547<\chi /\tilde{g}^2 <0$, whereas the Keck observations yield $-0.052157<\chi/\tilde{g}^2<0$. Both observationally allowed intervals are entirely contained within the stable region. Hence, the values of $\chi$ compatible with the EHT measurements correspond to stable configurations of Branch I.

It is instructive to compare this result with the standard EYM case.  Both theories admit exact non-Abelian RN solutions via the Wu-Yang monopole ansatz, and share the same background spacetime structure.
However, at the perturbative level, the induced global charge, generated by the quartic self-interaction, modifies the vector field dynamics in a way that stabilizes the hair without introducing higher derivatives, negative cosmological
constant, or any departure from the minimal coupling framework. 

This result has a natural interpretation within Bizon's conjecture \cite{Bizon:1994dh}, which posits that stable stationary BHs in any given  model are uniquely determined by their global charges. The EYM colored BHs are well-known counterexamples to the classical no-hair theorem but are also perturbatively unstable, and they carry additional non-global labels, most prominently the node number of the vector-field amplitude, so they do not test the generalized conjecture in its sharpest form. By contrast, the solutions of the present theory are uniquely parameterized by the two global charges $(M, Q_{\rm NA})$, with no additional discrete labels. This constitutes an explicit realization of Bizon's generalized no-hair conjecture for non-Abelian fields in $d=4$ asymptotically flat GR, the same regime in which the standard EYM colored BHs or non-Abelian Proca BHs fail to satisfy. The quartic self-interaction in Eq.~\eqref{eq:action} is, therefore, the simplest known modification of Yang-Mills theory that brings the non-Abelian hair into compliance with the generalized no-hair conjecture.
 
The present result establishes branch I as a potentially viable hairy BH solution, and opens a natural research program. The stability proof presented here is restricted to the spherically symmetric ($\ell = 0$) sector, in which the metric perturbations are non-dynamical and the analysis reduces to two scalar Schr\"{o}dinger-like problems for the vector-field modes. Extension to non-radial gravitational perturbations ($\ell \geq 2$), where the propagating metric degrees of freedom couple to the vector sector, is the natural next step and would yield the QNM spectrum of these BHs, providing one additional observational probe on the self-interaction coupling $\chi$ through the ringdown signal. The dynamical evolution of finite-amplitude perturbations, building on the well-posed scheme of \cite{Gomez:2025pag}, is a complementary direction that would further characterize the non-linear behavior of these solutions.

Finally, the theory has already shown remarkable astrophysical richness beyond the BH sector.  Particle-like soliton solutions, generalizing the Bartnik-McKinnon solutions \cite{1988PhRvL..61..141B}, were found in~\cite{Martinez:2022wsy}, while neutron-star configurations based on the regular 't~Hooft-Polyakov monopole were constructed in~\cite{Martinez:2024gsj}. Extending the present analysis to BH supported by the 't~Hooft-Polyakov or dyon configuration would complete the stability picture of the theory. More broadly, the existence of stable non-Abelian BHs opens a new avenue for the systematic study of their phenomenology and observational signatures.

\paragraph*{Acknowledgments.} 
We are grateful to Jorge A. Rueda for reading the original manuscript and giving useful suggestions. We thank the anonymous referees for their valuable comments and suggestions. J. F. R-R has been funded by Universidad Antonio Nari\~no under grant VCTI
2232026. 
\bibliography{biblio}

@article{Leaver:1990zz,
    author = "Leaver, Edward W.",
    title = "{Quasinormal modes of Reissner-Nordstrom black holes}",
    doi = "10.1103/PhysRevD.41.2986",
    journal = "Phys. Rev. D",
    volume = "41",
    pages = "2986--2997",
    year = "1990"
}

@article{Leaver:1985ax,
    author = "Leaver, E. W.",
    title = "{An Analytic representation for the quasi normal modes of Kerr black holes}",
    doi = "10.1098/rspa.1985.0119",
    journal = "Proc. Roy. Soc. Lond. A",
    volume = "402",
    pages = "285--298",
    year = "1985"
}

@misc{xAct,
  author       = {Martín-García, José M.},
  title        = {{xAct}: Efficient tensor computer algebra for the {Wolfram Language}},
  howpublished = {\url{http://www.xact.es/}}
}

@article{Witten:1976ck,
    author = "Witten, Edward",
    editor = "Shifman, Mikhail A.",
    title = "{Some Exact Multi - Instanton Solutions of Classical Yang-Mills Theory}",
    reportNumber = "HUTP-76/A172b",
    doi = "10.1103/PhysRevLett.38.121",
    journal = "Phys. Rev. Lett.",
    volume = "38",
    pages = "121--124",
    year = "1977"
}

@article{Gomez:2025pag,
    author = "Gomez, Gabriel and Rodriguez, Jose F.",
    title = "{Internal symmetry to the rescue: well-posed 1 + 1 evolution of self-interacting vector fields}",
    eprint = "2503.09757",
    archivePrefix = "arXiv",
    primaryClass = "gr-qc",
    doi = "10.1140/epjc/s10052-025-14657-1",
    journal = "Eur. Phys. J. C",
    volume = "85",
    number = "8",
    pages = "921",
    year = "2025"
}

@article{Bizon:1994dh,
    author = "Bizon, Piotr",
    title = "{Gravitating solitons and hairy black holes}",
    eprint = "gr-qc/9402016",
    archivePrefix = "arXiv",
    reportNumber = "UWTHPH-1994-5",
    journal = "Acta Phys. Polon. B",
    volume = "25",
    pages = "877--898",
    year = "1994"
}

@article{Ponglertsakul:2016fxj,
    author = "Ponglertsakul, Supakchai and Winstanley, Elizabeth",
    title = "{Solitons and hairy black holes in Einstein{\textendash}non-Abelian{\textendash}Proca theory in anti{\textendash}de Sitter spacetime}",
    eprint = "1606.04644",
    archivePrefix = "arXiv",
    primaryClass = "gr-qc",
    doi = "10.1103/PhysRevD.94.044048",
    journal = "Phys. Rev. D",
    volume = "94",
    number = "4",
    pages = "044048",
    year = "2016"
}

@article{Greene:1992fw,
    author = "Greene, Brian R. and Mathur, Samir D. and O'Neill, Christopher M.",
    title = "{Eluding the no hair conjecture: Black holes in spontaneously broken gauge theories}",
    eprint = "hep-th/9211007",
    archivePrefix = "arXiv",
    reportNumber = "CLNS-92-1162, MIT-CTP-2160",
    doi = "10.1103/PhysRevD.47.2242",
    journal = "Phys. Rev. D",
    volume = "47",
    pages = "2242--2259",
    year = "1993"
}

@article{Wald:1979lth,
    author = "Wald, Robert M.",
    title = "{Note on the stability of the Schwarzschild metric}",
    doi = "10.1063/1.524181",
    journal = "J. Math. Phys.",
    volume = "20",
    number = "6",
    pages = "1056",
    year = "1979"
}

@article{Robinson:1975bv,
    author = "Robinson, D. C.",
    title = "{Uniqueness of the Kerr black hole}",
    doi = "10.1103/PhysRevLett.34.905",
    journal = "Phys. Rev. Lett.",
    volume = "34",
    pages = "905--906",
    year = "1975"
}

@article{Martinez:2024gsj,
    author = "Martinez, Jhan N. and Rodriguez, Jose F. and Becerra, Laura M. and Rodriguez, Yeinzon and Gomez, Gabriel",
    title = "{Neutron stars in the generalized SU(2) Proca theory}",
    eprint = "2408.07674",
    archivePrefix = "arXiv",
    primaryClass = "gr-qc",
    reportNumber = "PI/UAN-2024-727FT",
    doi = "10.1103/PhysRevD.110.104070",
    journal = "Phys. Rev. D",
    volume = "110",
    number = "10",
    pages = "104070",
    year = "2024"
}

@article{Winstanley:1998sn,
    author = "Winstanley, E.",
    title = "{Existence of stable hairy black holes in SU(2) Einstein Yang-Mills theory with a negative cosmological constant}",
    eprint = "gr-qc/9812064",
    archivePrefix = "arXiv",
    reportNumber = "OUTP-98-92-P",
    doi = "10.1088/0264-9381/16/6/325",
    journal = "Class. Quant. Grav.",
    volume = "16",
    pages = "1963--1978",
    year = "1999"
}

@article{Carter:1971zc,
    author = "Carter, B.",
    title = "{Axisymmetric Black Hole Has Only Two Degrees of Freedom}",
    doi = "10.1103/PhysRevLett.26.331",
    journal = "Phys. Rev. Lett.",
    volume = "26",
    pages = "331--333",
    year = "1971"
}

@article{Israel:1967wq,
    author = "Israel, Werner",
    title = "{Event horizons in static vacuum space-times}",
    doi = "10.1103/PhysRev.164.1776",
    journal = "Phys. Rev.",
    volume = "164",
    pages = "1776--1779",
    year = "1967"
}

@article{Straumann:1990as,
    author = "Straumann, N. and Zhou, Z. H.",
    title = "{Instability of a colored black hole solution}",
    doi = "10.1016/0370-2693(90)90951-2",
    journal = "Phys. Lett. B",
    volume = "243",
    pages = "33--35",
    year = "1990"
}

@article{Allys:2016kbq,
    author = "Allys, Erwan and Peter, Patrick and Rodriguez, Yeinzon",
    title = "{Generalized SU(2) Proca Theory}",
    eprint = "1609.05870",
    archivePrefix = "arXiv",
    primaryClass = "hep-th",
    reportNumber = "PI-UAN-2016-597FT",
    doi = "10.1103/PhysRevD.94.084041",
    journal = "Phys. Rev. D",
    volume = "94",
    number = "8",
    pages = "084041",
    year = "2016"
}

@ARTICLE{1988PhRvL..61..141B,
       author = {{Bartnik}, Robert and {McKinnon}, John},
        title = "{Particlelike solutions of the Einstein-Yang-Mills equations}",
      journal = {\prl},
         year = 1988,
        month = jul,
       volume = {61},
       number = {2},
        pages = {141-144},
          doi = {10.1103/PhysRevLett.61.141},
       adsurl = {https://ui.adsabs.harvard.edu/abs/1988PhRvL..61..141B}
}

@article{Gomez:2023wei,
    author = "G{\'o}mez, Gabriel and Rodr{\'\i}guez, Jos{\'e} F.",
    title = {{New non-Abelian Reissner-Nordstr{\"o}m black hole solutions in the generalized SU(2) Proca theory and some astrophysical implications}},
    eprint = "2301.05222",
    archivePrefix = "arXiv",
    primaryClass = "gr-qc",
    doi = "10.1103/PhysRevD.108.024069",
    journal = "Phys. Rev. D",
    volume = "108",
    number = "2",
    pages = "024069",
    year = "2023"
}

@article{Martinez:2022wsy,
    author = "Martinez, Jhan N. and Rodriguez, Jose F. and Rodriguez, Yeinzon and Gomez, Gabriel",
    title = "{Particle-like solutions in the generalized SU(2) Proca theory}",
    eprint = "2212.13832",
    archivePrefix = "arXiv",
    primaryClass = "gr-qc",
    reportNumber = "PI/UAN-2023-723FT",
    doi = "10.1088/1475-7516/2023/04/032",
    journal = "JCAP",
    volume = "04",
    pages = "032",
    year = "2023"
}

@article{Radu:2011ip,
    author = "Radu, Eugen and Tchrakian, D. H.",
    title = "{Stable black hole solutions with non-Abelian fields}",
    eprint = "1111.0418",
    archivePrefix = "arXiv",
    primaryClass = "gr-qc",
    doi = "10.1103/PhysRevD.85.084022",
    journal = "Phys. Rev. D",
    volume = "85",
    pages = "084022",
    year = "2012"
}

@article{GallegoCadavid:2020dho,
    author = "Gallego Cadavid, Alexander and Rodriguez, Yeinzon and G\'omez, L. Gabriel",
    title = "{Generalized SU(2) Proca theory reconstructed and beyond}",
    eprint = "2009.03241",
    archivePrefix = "arXiv",
    primaryClass = "hep-th",
    reportNumber = "PI/UAN-2020-678FT",
    doi = "10.1103/PhysRevD.102.104066",
    journal = "Phys. Rev. D",
    volume = "102",
    number = "10",
    pages = "104066",
    year = "2020"
}

@article{Volkov:1998cc,
    author = "Volkov, Mikhail S. and Gal'tsov, Dmitri V.",
    title = "{Gravitating nonAbelian solitons and black holes with Yang-Mills fields}",
    eprint = "hep-th/9810070",
    archivePrefix = "arXiv",
    reportNumber = "ZU-TH-17-98, YITP-98-53, DTP-MSU-31-98",
    doi = "10.1016/S0370-1573(99)00010-1",
    journal = "Phys. Rept.",
    volume = "319",
    pages = "1--83",
    year = "1999"
}

@article{Bizon:1990sr,
    author = "Bizon, P.",
    title = "{Colored black holes}",
    doi = "10.1103/PhysRevLett.64.2844",
    journal = "Phys. Rev. Lett.",
    volume = "64",
    pages = "2844--2847",
    year = "1990"
}

@article{Volkov:1989fi,
    author = "Volkov, M. S. and Galtsov, D. V.",
    title = "{NonAbelian Einstein Yang-Mills black holes}",
    journal = "JETP Lett.",
    volume = "50",
    pages = "346--350",
    year = "1989"
}

@article{Lee:1991qs,
    author = "Lee, Ki-Myeong and Nair, V. P. and Weinberg, Erick J.",
    title = "{A Classical instability of Reissner-Nordstrom solutions and the fate of magnetically charged black holes}",
    eprint = "hep-th/9111045",
    archivePrefix = "arXiv",
    reportNumber = "CU-TP-540, FERMILAB-PUB-91-326-A",
    doi = "10.1103/PhysRevLett.68.1100",
    journal = "Phys. Rev. Lett.",
    volume = "68",
    pages = "1100--1103",
    year = "1992"
}

@article{Horndeski:1974wa,
    author = "Horndeski, Gregory Walter",
    title = "{Second-order scalar-tensor field equations in a four-dimensional space}",
    doi = "10.1007/BF01807638",
    journal = "Int. J. Theor. Phys.",
    volume = "10",
    pages = "363--384",
    year = "1974"
}
\end{document}